\documentclass[a4paper,11pt]{article}
\usepackage[utf8]{inputenc}
\usepackage{newtxtext}
\usepackage{cite}
\usepackage{physics}
\usepackage{graphicx}
\usepackage{amsmath}
\usepackage{amssymb}
\usepackage[font=small,skip=1pt]{caption}
\usepackage[rmargin = 2.5cm, lmargin = 2.5cm, tmargin = 2.5cm, bmargin = 2.5cm]{geometry}
\usepackage{multicol}
\usepackage{float}
\usepackage{authblk}
\usepackage{hyperref}

\hypersetup{
    colorlinks=true,
    citecolor=blue,
    linkcolor=blue,
    filecolor=magenta,      
    urlcolor=blue,
}

\title{{Optimal allocation of finite sampling capacity in accumulator models of multi-alternative decision making}\vspace{10pt} 
}

\author[1]{Jorge Ramírez-Ruiz}
\author[1,2]{Rubén Moreno-Bote}
\affil[1]{Center for Brain and Cognition, and Department of Information and Communication Technologies, Universitat Pompeu Fabra, Barcelona, Spain}
\affil[2]{Serra Húnter Fellow Programme, Universitat Pompeu Fabra, Barcelona, Spain}
\date{}

\begin{document}

\maketitle

\section*{Abstract}
When facing many options, 
we narrow down our focus to very few of them. 
Although behaviors like this can be a sign of heuristics, they can actually be optimal under limited cognitive resources. Here we study the problem of how to optimally allocate limited sampling time to multiple options, modelled as accumulators of noisy evidence, to determine the most profitable one.
We show that the effective sampling capacity of an agent increases with both available time and the discriminability of the options, and optimal policies
undergo a sharp transition as a function of it. For small capacity, it is best to allocate time evenly to exactly five options and to ignore all the others, regardless of the prior distribution of rewards.
For large capacities, the optimal number of sampled accumulators grows sub-linearly, closely following a power law for a wide variety of priors. 
We find that allocating equal times to the sampled accumulators is better than using uneven time allocations.
Our work highlights that multi-alternative decisions are endowed with breadth-depth tradeoffs, demonstrates how their optimal solutions depend on the amount of limited resources and the variability of the environment, and shows
that narrowing down to a handful of options is always optimal for small capacities.


\section{Introduction}

The problem of allocating finite resources to determine the best of several options is common in decision making, from deciding which vaccine candidates to fund for further research to choosing a movie for Saturday night. 
In these cases, planning, and thus resource allocation, needs to be made in advance, well before feedback about the success of the choice is observed. Consequently, two important questions arise: How many options should we examine? And, for how long? When resources are limited, such as number of participants or weekend free time, a decision maker should balance breadth, how many options to sample, and depth, how much to sample each. This ubiquitous decision making problem under constrained resources is what has been called the breadth-depth (BD) dilemma \cite{miller1981depth, horowitz1978depth,BDpnas}.

In the face of many alternatives, humans quickly narrow down the number of considered options to around two to five \cite{payne1976task,olshavsky1979task,beach1993broadening,levin1998choosing,hauser1990considerationset}, and, when presented with more than 6 options, experienced overload produces suboptimal choices in certain conditions \cite{iyengar2000choice,scheibehenne2010can}. 
Models describe this behavior by assuming that considering more options incurs search or mental costs \cite{hauser1990considerationset,mehta2003price,stigler1961economics}, but why people consider small sets in a wide range of environments is still a matter of debate.
While this could be explained by strict small capacity limits in attention or working memory \cite{miller1956magical,cowan2005capacity}, the nature of this small capacity would still need to be addressed \cite{brady2016WM}. Another possibility is that capacity is not necessarily small, but rather that sampling few options and ignoring the vast majority is actually an optimal policy that favors depth over breadth \cite{BDpnas}. This possibility is supported by the fact that neuronal resources devoted to decision making are not precisely low, as dozens of brain areas and several billions of neurons are involved in even simple decision making tasks \cite{rushworth2011frontal,siegel2015cortical,vickery2011ubiquity,yoo2018economic}. Thus, processing bottlenecks could be reflections of close-to-optimal policies. 

Bounded rationality accounts \cite{simon1972theories,russell1991principles,gershman2015computational,griffiths2015rational} surmise that many features of cognition arise from the finite limits of the nervous system. 
This must also be the case for the nature of the policies chosen by people in decision making, but oftentimes the constraints imposed by the limited resources are not made explicit. Indeed, choices between two or three options have been typically modelled as optimal stopping problems \cite{ratcliff1976retrieval,gold2007neural,krajbich-rangel2011multialternative,drugowitsch2012cost,tajima2019optimal,jang-drugowitsch2020optimal,callaway2020fixation}, where agents should optimally balance the prospect of learning the value of the options with the costs of sampling them, but they do so without computational or capacity constraints. The effect of resource limitations on decision making might not be important when there are only two or three available options, but it might be critical when going beyond those low numbers. In that case, the allocation of resources might be governed by two-stage processes \cite{hauser1990considerationset,mehta2003price,shocker1991consideration,roberts1991development}, instead of purely sequential processes, where the first decision is about the subset of options that will be considered for further processing. 

Here we study whether narrowing attention to a few options results from optimally allocating finite resources. To this end, we consider an infinitely divisible sampling resource (e.g. time), such that there are no bounds in the number of alternatives that can be considered. In our model, an agent can first allocate finite sampling time over an arbitrarily large number of options, modelled as accumulators of noisy evidence, with the only restriction that the total sampling time is fixed.
Accumulation of evidence runs in parallel and independently for each accumulator, and only their final states are observed. 
Based on the observations, the agent picks up the one with the highest expected drift rate, which defines the utility of the choice. The goal of the agent is to optimize the allocation of sampling time such that expected utility is maximized.
We identify a critical variable in the problem, that we simply call \emph{capacity}, that increases with the actual size of the resources of the agent as well as with the discriminability between options, and we find that this capacity separates two distinct regimes of optimal allocation.
When sampling capacity is small, the optimal policy is to sample exactly five options, regardless of the prior. In contrast, when capacity is large, the number of options to sample grows with capacity in a sub-linear fashion that depends on the prior.
We find a duality between allocated time and allocated precision to the options, such that all our results generalize to allocating precision while keeping fixed sampling time.
Finally, we show that even allocations are optimal, and thus better than more complex asymmetric time allocations over the considered options. Overall, our results suggest that decisional bottlenecks can be a byproduct of optimal policies in the face of uncertainty.

\section{Results}
 \subsection*{Multi-accumulator model}
 
We consider an environment that generates many options ($N \gg 1$) from which to choose (Fig. 1, top), each one characterized by a `drift' parameter $\mu_i$ ($i=1,\ldots,N$), unknown to the agent. All drifts $\mu_i$ are drawn identically and independently from a prior probability distribution $p_\theta(\mu)$, known to the agent and assumed to have finite mean and variance. 

The agent learns about the options, in order to choose between them, by allocating sampling time $t_i$ to each (Fig. 1, bottom). The critical aspect of our model is that sampling times $t_i \geq 0$ need to be allocated before feedback about the drifts is received, and with the constraint that the total sampling time equals a finite available time $T$,
\begin{equation}
    \sum_{i=1}^{N} t_i = T.
    \label{eq:T-constraint}
\end{equation}
In practice, the agent needs to decide on the number of options $M \leq N$ to be sampled and their corresponding sampling times $t_i>0$ for $i \leq M$, while the remaining options $i>M$ are ignored by giving them no sampling time, $t_i=0$. The ordering of the options is irrelevant, as they are initially indistinguishable, and thus we take the first $M$ as those that are sampled. We assume that non-sampled options cannot be chosen, although a `default' option can be added to our framework with no change of our main results. 

Once total sampling time is allocated, noisy evidence about the drift $\mu_i$ of each of the sampled options $i \leq M$ is integrated by independent accumulators (Fig. 1, middle) according to the drift-diffusion process
\begin{equation}\label{eq:ddm}
    \frac{d x_i(t)}{dt}=\mu_i+\eta_i\left(t\right),
\end{equation}
where $x_i(t)$ is the accumulated evidence up to time $t$ with initial condition $x_i(0)=0$, and $\eta_i\left(t\right)$ is a Gaussian white noise with zero mean and fixed variance $\sigma^2$, independent and identical for all the accumulators.

\begin{figure}[t]
\center

\includegraphics[width=0.9\textwidth]{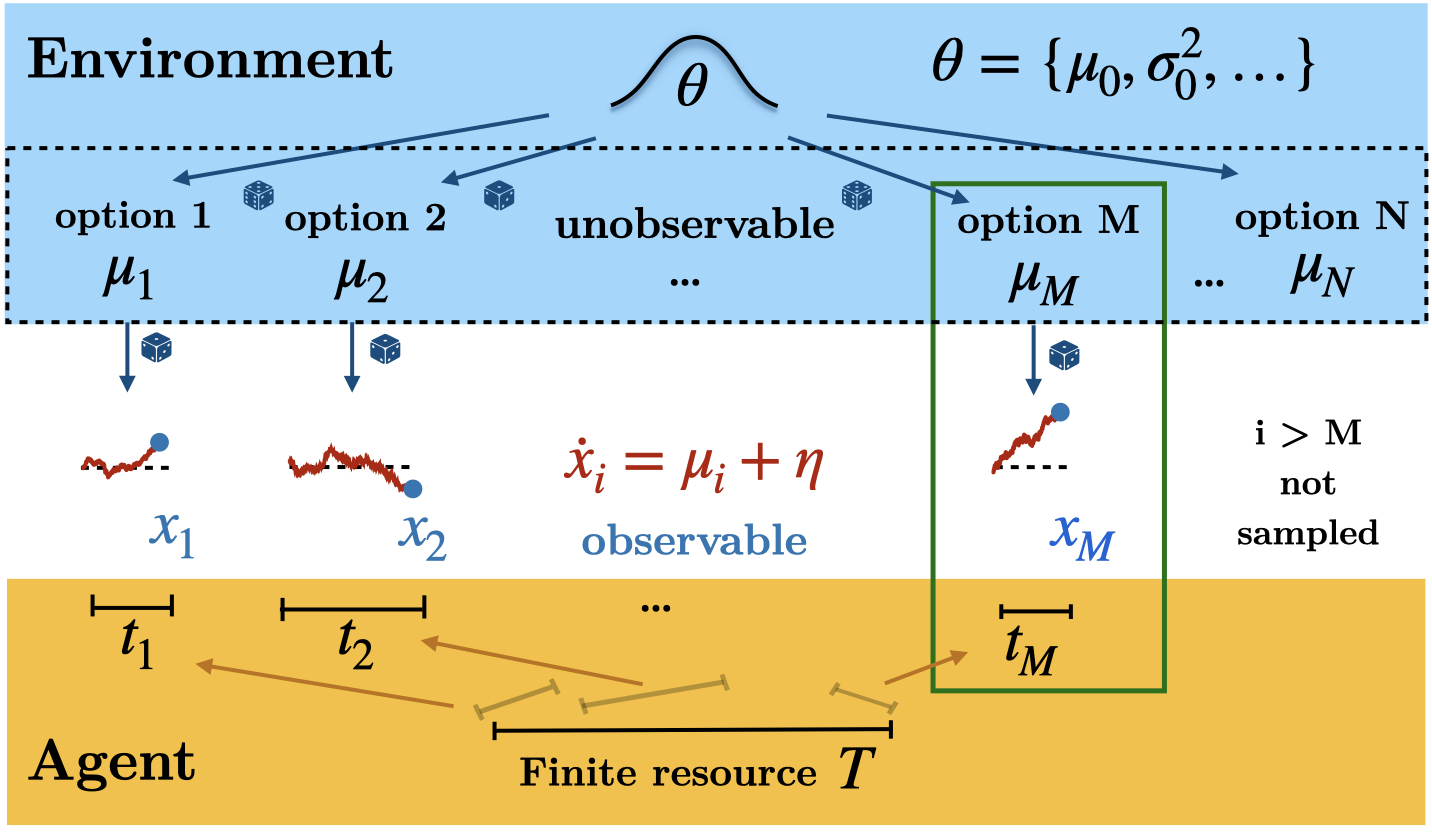} \newline
\caption{A multi-accumulator model with finite sampling resources.
The environment produces a large number of options, each characterized by a drift $\mu_i$, unknown to the agent and drawn from a prior distribution characterized by hyperparameters $\theta$, which is known to the agent. The agent has a finite resource $T$, that they divide and allocate across options, $\sum_it_i=T$, in order to sample them. In practice, the agent allocates finite sampling time to a finite number $M$ of accumulators to infer their unknown drifts. After allocation, evidence (red lines) is optimally integrated by the accumulators. The agent observes the integrated evidence $x_i$ at the allocated time $t_i$, infers the drifts for each of the accumulators and chooses the one that is deemed to have the highest drift (in this case, $\mu_M$; green box).
}
\label{fig:fig1}
\end{figure}

The result of the accumulation is the total evidence $x_i$ at time $t_i$, both of which are observed by the agent and constitute the sufficient statistics for the unknown drift $\mu_i$ \cite{moreno2010decision}. 
With these observations, the agent builds the posterior distribution of the drifts by using Bayes rule as
\begin{equation}
    p(\mu_i | x_i, t_i, \sigma, \theta) 
    = \frac{\mathcal{L}(\mu_i | x_i,t_i,\sigma) p_\theta(\mu )}{p(x_i | t_i,\sigma,\theta)}\label{eq:bayes},
    \vspace{5pt}
\end{equation}
where $\mathcal{L}(\mu_i | x_i,t_i,\sigma) = \mathcal{N}(x_i | \mu_i t_i, \sigma^2 t_i)$ is the likelihood function for the drift, 
$p_\theta(\mu)$ is the prior distribution
and $p(x_i | t_i,\sigma,\theta) = \int \dd{\mu} \mathcal{N}(x_i | \mu t_i, \sigma^2 t_i)p_\theta(\mu)$
is the marginal distribution of the evidence, which serves as a normalization constant.
The posterior mean of the drift becomes 
$\hat{\mu}_i \equiv \hat{\mu}_i\left(x_i,t_i,\sigma,\theta\right) \equiv \int \dd{\mu} \mu\, p(\mu | x_i, t_i, \sigma, \theta)$, which is finally used by the agent to choose the option with the highest inferred drift (Fig. 1, middle, green box), 
to optimize utility,  $U\left(M,\mathbf{x},\mathbf{t},\sigma,\theta\right)=\max_{i \leq M}
\hat{\mu}_i\left(x_i,t_i,\sigma,\theta\right)$, where $\mathbf{x}=(x_1,...,x_M)$ is the vector of observations for the $M$ accumulators with allocated times $\mathbf{t}=(t_1,...,t_M)$. To avoid notation clutter, from now on we will stop writing the dependence on $\sigma$ and $\theta$ of the various functions and leave it implicit. 

The last expression is the utility of the choice of accumulator, which depends on the observations and allocation times. However, before time is allocated, the observations $\mathbf{x}$ themselves will be unknown to the agent. Therefore, the expected utility of a given allocation $\mathbf{t}$ is given by taking the expectation of the above utility over all possible observations as
 \begin{equation}\label{eq:utility_a} 
    \hat{U}(M,\mathbf{t}) \equiv \mathbb{E}\left[\max_{i \leq M}\hat{\mu}_i | \mathbf{t}
    \right]
    = \int \dd{x_1} ... \dd{x_M} p(x_1, ..., x_M | \mathbf{t})\max_i\hat{\mu}_i\left(x_i,t_i\right) , 
\end{equation}
where, using the independence of the accumulators, $p(x_1, ..., x_M | \mathbf{t}) = \prod_{k \leq M} p(x_k|t_k)$ is the product of the marginal distribution of the evidences. 

Optimally inferring the drifts from observations is readily accessible through Bayesian inference as shown above. Thus, the main, and harder, objective of the agent is to optimize the allocation policy, i.e. to select both the number of sampled accumulators $M \leq N$ and the time $t_i$ allocated to each, in order to maximize expected reward, while satisfying the total sampling time constraint in Eq. (\ref{eq:T-constraint}). This is accomplished by optimizing the utility with respect to $M$ and $\mathbf{t}=(t_1,...,t_M)$ as
\begin{equation}\label{eq:utility_t} 
    (M^*,\mathbf{t}^*) = \arg\max_{M,\;\mathbf{t}} \hat{U}(M,\mathbf{t}) .
\end{equation}

 \subsection*{Capacity and time-precision duality}
 
\begin{figure}[t]
\center

\includegraphics[width=0.8\textwidth]{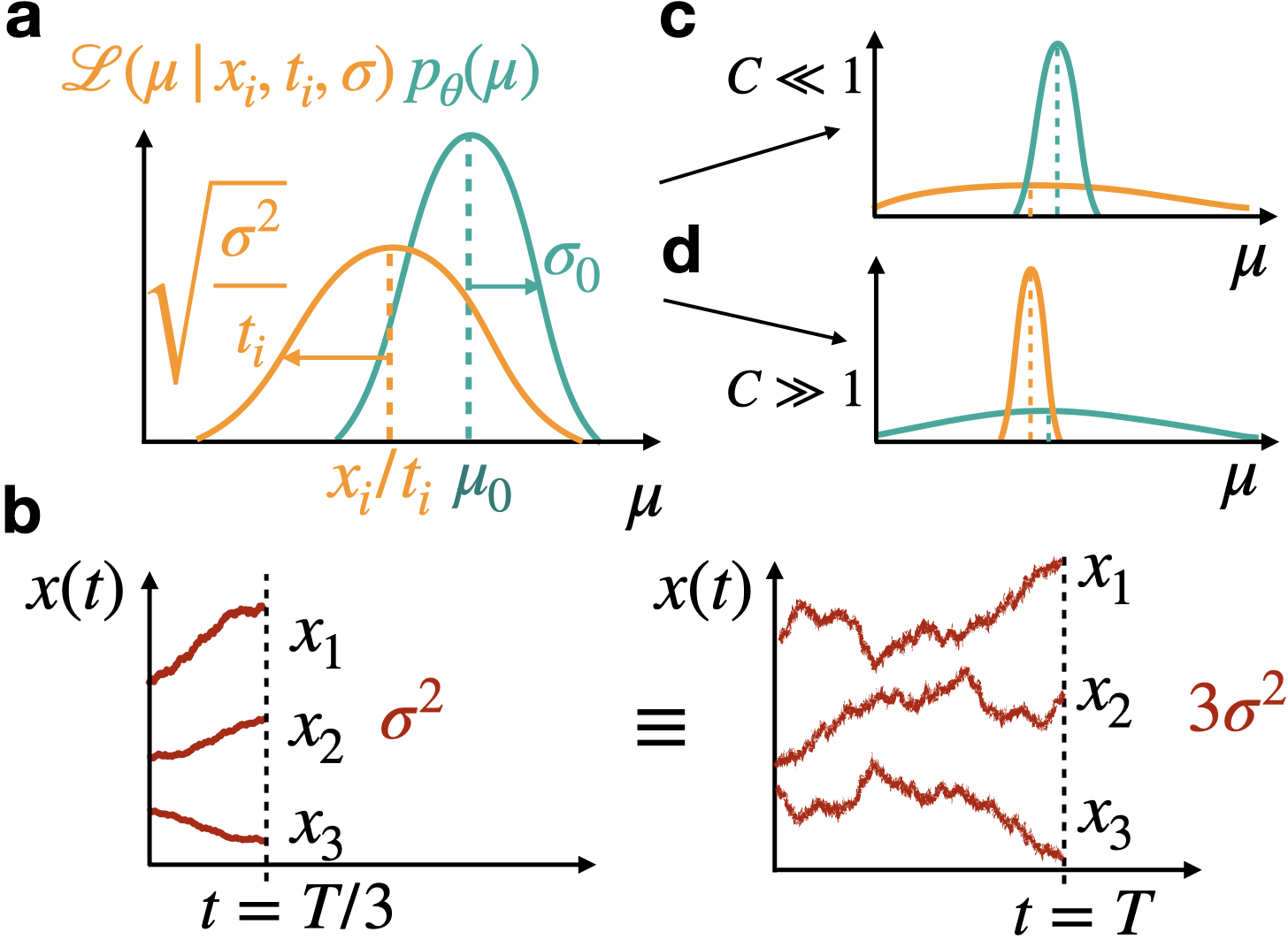} \newline
 \caption{Time/precision duality and the notion of capacity. (\textbf{a}) The likelihood of the drift $\mu$ (in orange) given the evidence has variance $\sigma^2/t_i$ and the prior distribution of the drifts (in cyan) has variance $\sigma_0^2$. These quantities determine capacity as in Eq. (\ref{eq:definition_capacity}). (\textbf{b}) Time and sampling noise are intricately related (see text). In this example, allocating time $T/3$ to each accumulator under fixed precision $1/\sigma^2$ (left) is equivalent to allocating precision $1/3\sigma^2$ to each accumulator under fixed sampling time $T$ (right). (\textbf{c}) Small capacity means that the variance of the observation is much larger than the variance of the prior, indicating that it is difficult to confidently identify the best drift from the observations. (\textbf{d}) In the large capacity limit, it is easier to differentiate the good drifts from the poor ones.}
 
  \label{fig:fig2}
 \end{figure}

While time is the resource that the agent allocates, we found a dimensionless scale that expresses their actual sampling capacity, i.e. their ability to sample and differentiate between drifts, which we call capacity $C$ (Fig. \ref{fig:fig2}).
As the agent integrates noisy evidence through Eq. (\ref{eq:ddm}), the likelihood of the drift $\mu_i$ for accumulator $i$ is proportional to a Gaussian (Fig. \ref{fig:fig2}\textbf{a}, orange curve) with mean $x_i/t_i$ and variance $\sigma^2/t_i$, $\mathcal{L}(\mu_i | x_i,t_i,\sigma) \propto \mathcal{N}(\mu_i | \frac{x_i}{t_i} \frac{\sigma^2}{t_i})$ \cite{moreno2010decision}. Its variance $\sigma^2/t_i$ shows how the sampling time and the variance of the sampling noise are related when inferring the drift $\mu_i$, since the Gaussian gets broader with increasing $\sigma$ or decreasing time $t_i$. In fact, the sampling capacity of the agent should capture this duality. Thus, having a fixed capacity could be interpreted as having a fixed noise variance $\sigma^2$ for all accumulators and allocating time $T$ between them (Fig. \ref{fig:fig2}\textbf{b}, left) or as having a fixed sampling time $T$ for each of the accumulators and allocating precision $1/\sigma^2$ between them (Fig. \ref{fig:fig2}\textbf{b}, right).

Moreover, the posterior in Eq. (\ref{eq:bayes}) depends on the prior as well (Fig. \ref{fig:fig2}\textbf{a}, cyan curve). For fixed evidence, the broader the prior is, the easier it is to differentiate between sampled drifts, since the expected squared distance between two drifts drawn from the same distribution is twice its variance ${\rm Var}[p_\theta(\mu)]$. Therefore, we define the capacity allocated to option $i$ as the ratio between the precision of the observation and the precision of the prior,
\begin{equation}\label{eq:definition_capacity}
    c_i = \frac{{\rm Var}[p_\theta(\mu)]}{{\rm Var}[\mathcal{N}(\mu_i | \frac{x_i}{t_i} , \frac{\sigma^2}{t_i})]} = \frac{\sigma_0^2}{\sigma^2} t_i.
\end{equation}
Adding the individual capacities results in the total sampling capacity of the agent,
\begin{equation}\label{eq:total_capacity}
    C = \sum_i c_i
    = \frac{\sigma_0^2}{\sigma^2}T.
\end{equation}

For the rest of this article, we stick to the interpretation of allocating capacity as dividing the total time $T$ while fixing the accumulation noise $\sigma$, such that the variable we can control is the sampling time allocated to each option, keeping in mind that all the results presented below can be readily reinterpreted as dividing precision while giving to all options the same sampling time.

\subsection*{Even sampling}

Optimally dividing sampling capacity $C$ into options is an \textit{a priori} hard problem due to its high dimensionality. However, we show in a section below through numerical simulations and for Gaussian priors that the optimal allocation lies within the family of even allocations, where $M$ options receive equal sampling time $t_i=t \equiv T/M$, while the remaining others are given no time. Thus, finding the optimal policy reduces to finding the optimal number $M$ of accumulators to sample.

In this section, we exploit the structure of even sampling. First, the posterior mean of the drift $\hat{\mu}_i(x_i , t)$, computed from Eq. (\ref{eq:bayes}), is a monotonously increasing function of the evidence $x_i$ for any prior (see proof in Sec. \ref{subsec:expectedmu} of the Methods). Therefore, the option that maximizes the posterior mean $\hat{\mu}_i$ is the one that has the highest evidence $x_i(t)$, as all $M$ sampled options are given the same sampling time $t$. This allows us to work by maximizing evidence instead of maximizing the posterior means of the drifts in Eq. (\ref{eq:utility_a}).
Secondly, by changing variables $y \equiv \max_i x_i$, and using the probability distribution of the maximum $y$, denoted $p_{{\rm max}}(y |t, \sigma, \theta)$, the expected utility in Eq. (\ref{eq:utility_a}) can be recast in the one-dimensional integral
\[
    \hat{U}(M,t) = \int\dd{y}\, p_{{\rm max}}(y |t)\hat{\mu}(y , t).
\]
Finally, given that the $M$ options are sampled evenly, the probability distribution of the maximum can be simplified by using the cumulative distribution of the evidence $x$ for an arbitrary accumulator, $F_x(y| t) = \int_{-\infty}^y\dd{x'}p(x'|t)$, where $p(x | t)$ is the marginal of the evidence $x$ of the accumulator, as 
\begin{equation}\label{eq:cum_max_iid}
    p_{\rm max}(y |t) = \dv{}{y}\left[F_x(y|t)\right]^M. 
\end{equation}
With all the above, the expected utility in Eq. (\ref{eq:utility_a}) can thus be written as
\begin{equation}\label{eq:utility_b}
    \hat{U}(M,t) = 
    M\int\dd{y}\, \left[F_x(y|t)\right]^{M-1} p(y|t )\,\hat{\mu}(y , t).
\end{equation}
When the prior distribution is a Gaussian with mean $\mu_0$ and variance $\sigma_0^2$, it is possible to identify the total capacity $C = \frac{\sigma_0^2}{\sigma^2}T$ explicitly and Eq. (\ref{eq:utility_b}) simplifies to
\begin{equation}\label{eq:utilityG}
    \hat{U}(M,C) 
    = \mu_0 + \frac{M\sigma_0}{\sqrt{1+\frac{M}{C}}}\int_{-\infty}^{\infty} \dd{y}  \left[\Phi(y)\right]^{M-1}\mathcal{N}(y|0,1)\,y ,
\end{equation}
where
$
\Phi(y)
= \frac{1}{2}\left[1 + \erf\left(\frac{y}{\sqrt{2}}\right)\right]
$
is the cumulative distribution function of a normal distribution. 
\vspace{0.5em}

\begin{figure}[t]
\center
\includegraphics[width=0.75\textwidth]{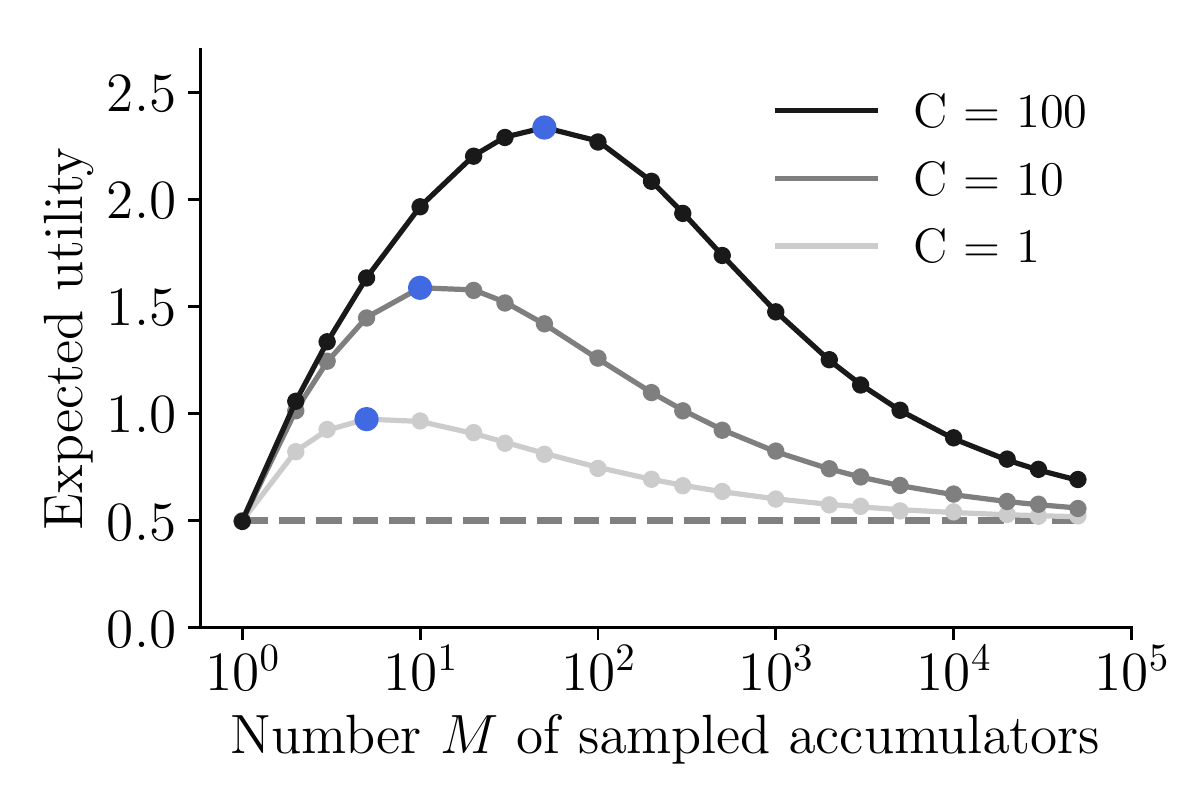}
\captionsetup{width=.75\linewidth}
\vspace{1em}
 \caption{
Expected utility as a function of sampled accumulators exhibits the breadth-depth tradeoff. Results for the Gaussian prior case ($\mu_0 = 0.5, \sigma_0^2 = 1$), for three different capacities. Blue points denote the maxima. Note log horizontal scale (points, Monte Carlo simulations; lines, theoretical predictions, Eq. \ref{eq:utilityG}).  
}
 \label{fig:fig3a}
 \end{figure}

Plotting the utility in Eq. (\ref{eq:utilityG}) as a function of the number of sampled accumulators $M$ reveals a clear breadth-depth tradeoff (Fig. \ref{fig:fig3a}). 
At the depth limit, $M = 1$, only one accumulator is sampled and it is given all sampling time $T$. In this case,  the expected utility will simply be the expected value of the prior, $\mu_0$ (Fig. \ref{fig:fig3a}, left point), since there is no choice to be made between accumulators. At the breadth extreme, $M/C \rightarrow \infty$, the evidence gathered for each accumulator is very noisy because each has been allocated a very short sampling time, and thus choosing any will amount to an expected utility again equal to the prior mean (rightmost points). Therefore, for all capacities, there is an intermediate optimal value for the number of accumulators to sample, $M^*$.

\subsection*{Sharp transition between the small and large capacity regimes}
Our main result is that the optimal allocation policies are qualitatively different at small and large capacity, and that there is a abrupt transition between the two regimes. We provide useful asymptotic analytical expressions for the utility in Eq. (\ref{eq:utility_b}) and the optimal $M^*$ in both limits and describe their characteristic features.
 
\begin{figure}[t]
\center
\includegraphics[width=0.9\textwidth]{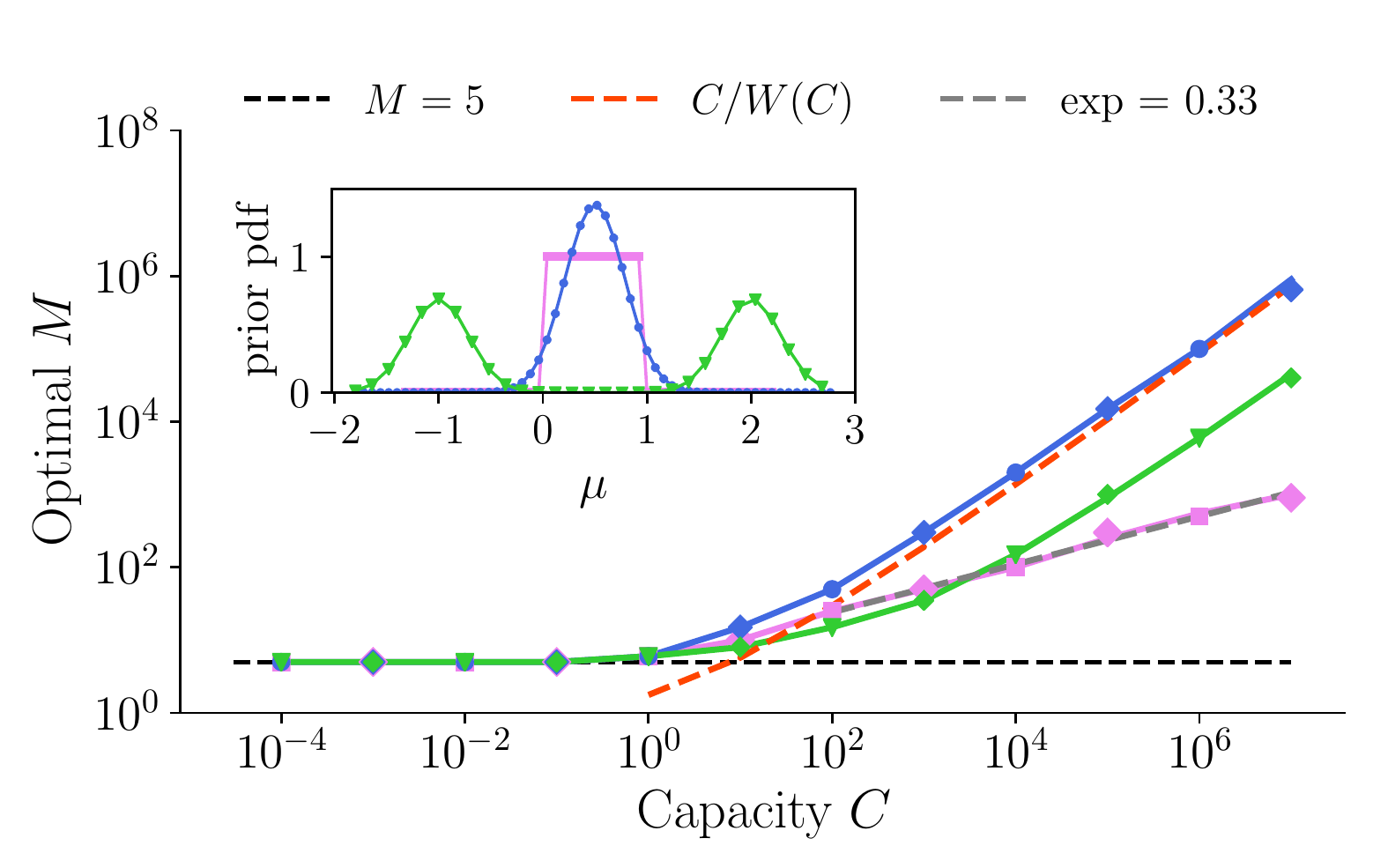}
\captionsetup{width=.9\linewidth}
\vspace{1em}
 \caption{
 The optimal number of sampled accumulators undergoes qualitatively different behaviors at small and large capacity values. Results come from searching the maximum expected utility via Monte Carlo simulations (points) and numerical integration (lines) for a Gaussian (Eq. \ref{eq:utilityG}), uniform (Eq. \ref{eq:utility_flat}) and a bimodal (Eq. \ref{eq:utility_bimodal}) priors (illustrated in inset). In addition, diamond markers indicate simulations with a `default' option. We used $\mu_0 = 0.5$ for all priors and $\sigma_0^2 = 1/12$ for the Gaussian prior to match the uniform distribution. For the bimodal prior, the variance of each mode equals $\sigma_0^2$. Dashed red line corresponds to the asymptotic limit in the Gaussian prior case, Eq. (\ref{eq:asymptoticM_Gaussian}). Dashed gray line is the best power law fit for the uniform prior case. 
}
 \label{fig:fig3}
 \end{figure}

The limit $C \ll 1$ corresponds to the case where the uncertainty in the observation $\sigma^2/T$ is much larger than the variance of the prior $\sigma_0^2$, i.e. the Gaussian likelihood is much wider than the prior (Fig. \ref{fig:fig2}\textbf{c}). In this limit, we find that the utility in Eq. (\ref{eq:utility_b}) can be expanded a series in powers of $\sqrt{C}$, which at first order is given by (see Sec. \ref{subsec:evalmu})
\begin{equation}\label{eq:utility_small_C}
    \hat{U}(M,C) = \mu_0 + \sigma_0 \sqrt{\frac{C}{2\pi}}\left[ \sqrt{M}\int_{-\infty}^\infty \dd{z} z \exp(-\frac{z^2}{2})\left(\frac{1}{2} + \frac{1}{2}\erf\left(\frac{z}{\sqrt{2}}\right)\right)^{M-1}\right] + \mathcal{O}(C).
\end{equation}
Remarkably, this expression holds for any prior distribution as long as capacity is small enough.
Using Extreme Value Theory (see Sec. \ref{subsec:asymptotic}) and noting that the only depencence of $M$ is in the quantity in the square brackets, we find that utility decreases with $M$ for large $M$.
On the other hand, it is easy to see that expected utility attains its lowest value when $M=1$. Thus, as utility is positive, the optimal $M$ should be attained at some intermediate value. We find numerically that the optimum happens when
\[
    M^* (C \ll 1) = 5.
\]
In summary, at small capacity the optimal number of sampled accumulators is constant and equal to five, regardless of the prior and the value of capacity. We have confirmed this strong prediction by direct numerical integration of Eq. (\ref{eq:utility_b}) using different prior distributions, including Gaussian, uniform and bimodal (Fig. \ref{fig:fig3}), which also holds even when a non-sampled, default, option can be chosen (diamond markers).

The opposite limit $C \gg 1$
corresponds to the case where the precision of the observation is much greater than the one of the prior (Fig. \ref{fig:fig2}\textbf{d}). Intuitively, this means that the quality of the observations is good enough to likely differentiate the drifts between two randomly chosen accumulators, and thus we expect that the optimal number of accumulators to increase with increasing capacity, giving a qualitatively different behaviour than the small capacity limit. 
To study this limit, we assume that the optimal number of sampled options $M^*$ increases with $C$, an assumption that is consistent with the results shown below, and thus we study the behaviour of Eq. (\ref{eq:utility_b}) for large $M$. In this limit, we can find an analytical expression for the expected utility in Eq. (\ref{eq:utilityG}), when the prior distribution is Gaussian,

\begin{equation}\label{eq:asymptoticGaussUtility}
    \hat{U}(M,C) \rightarrow \mu_0 + \sigma_0\frac{b_M}{\sqrt{1+\frac{M}{C}}},
\end{equation}
where $ b_M = \left(2 \log(M) - \log(\log(M)) - \log(4\pi)\right)^{1/2}$ (see Sec. \ref{subsec:asymptotic}).
By relaxing $M$ to be continuous, we can maximize expected utility, and we find that the optimal number of sampled options for large capacity satisfies, up to leading order, the implicit equation $M^* \log(M^*) = C$. After inverting it, the optimal number of sampled options is
\begin{equation}\label{eq:asymptoticM_Gaussian}
    M^*(C \gg 1) = \frac{C}{W(C)},
\end{equation}
where $W(C)$ is the Lambert function. 
This asymptotic limit provides a very good approximation to the optimal $M^*$ at large $C$ obtained from direct numerical integration of Eq. (\ref{eq:utilityG}) (Fig. \ref{fig:fig3}; red dashed line, theory; blue points, simulations). 
For prior distributions other than the Gaussian, we rely on numerical integration of Eq. (\ref{eq:utility_b}) (see Secs. \ref{subsec:uniform_prior} and \ref{subsec:bimodal} for analytical expressions).
For a uniform prior, the optimal number of sampled options increases as a power law with an exponent close to $1/3$ (Fig. \ref{fig:fig3}, pink), while for a bimodal prior the optimal number increases in a similar fashion to the Gaussian prior case (green). While differences of asymptotic limits are due to the presence of bounded or unbounded drifts in the priors, 
in all cases the increase is sub-linear, indicating that increasingly longer times are allocated to each of the sampled accumulators as capacity increases. 

The above results show that there are two distinct regimes, one at small and another at large capacities, characterized by qualitatively different optimal allocations: while at small capacity the optimal number of sampled options should be five regardless of the prior, at large capacity the optimal number of sampled options grows sublinearly regardless of the tested prior. 
Further, we observe that there is an abrupt transition between the two regimes as capacity grows, with a bump being observed at intermediate capacity values.

\subsection*{Even allocation is optimal}
Above we have assumed that we could find the optimal time allocation within the subset of even allocations, such that, given finite total time $T$, an agent just needs to determine how many options will be sampled and split equal time to all of them. Conveniently, this set is discrete and thus amenable to effective search of the optimum. However, in general, the set of allocation policies is the  infinite-dimensional simplex $\sum_i t_i = T$, $t_i \ge 0$ for all $i$, as a priori the agent could unevenly split time to options in any arbitrary way. Despite its infinite-dimensionality, we have seen in the case of even sampling that it is optimal to ignore (infinitely) many options, such that $t_i > 0$ only for $i \in \{1,...,M\}$, with finite $M$, and then we say that there are $M$ active dimensions. 

To address the most general case, using the above intuitions we first generalize the expected utility, Eq. (\ref{eq:utility_b}), to the case when allocated time is unevenly distributed among $M$ accumulators, as
\begin{equation}\label{eq:ut_arbitrary}
    \hat{U}(M,\mathbf{t}) = \int_{-\infty}^{\infty} \dd{y} \dv{}{y}\left[\prod_{i=1}^M F_x(y|t_i )\right] \hat{\mu}(y,t_i) ,
\end{equation}
where $F_x(y|t_i)$ is the cumulative distribution function of the posterior when using $t_i$ sampling time. Our goal is then, for every $M$, to find the allocation $\mathbf{t}$ that maximizes Eq. (\ref{eq:ut_arbitrary}) under the capacity equality constraint and the inequalities $t_i \ge 0$ for all $i$, and then select the optimal $M$, the one that achieves the highest utility.

\begin{figure}[t]
\center
\includegraphics[width=\textwidth]{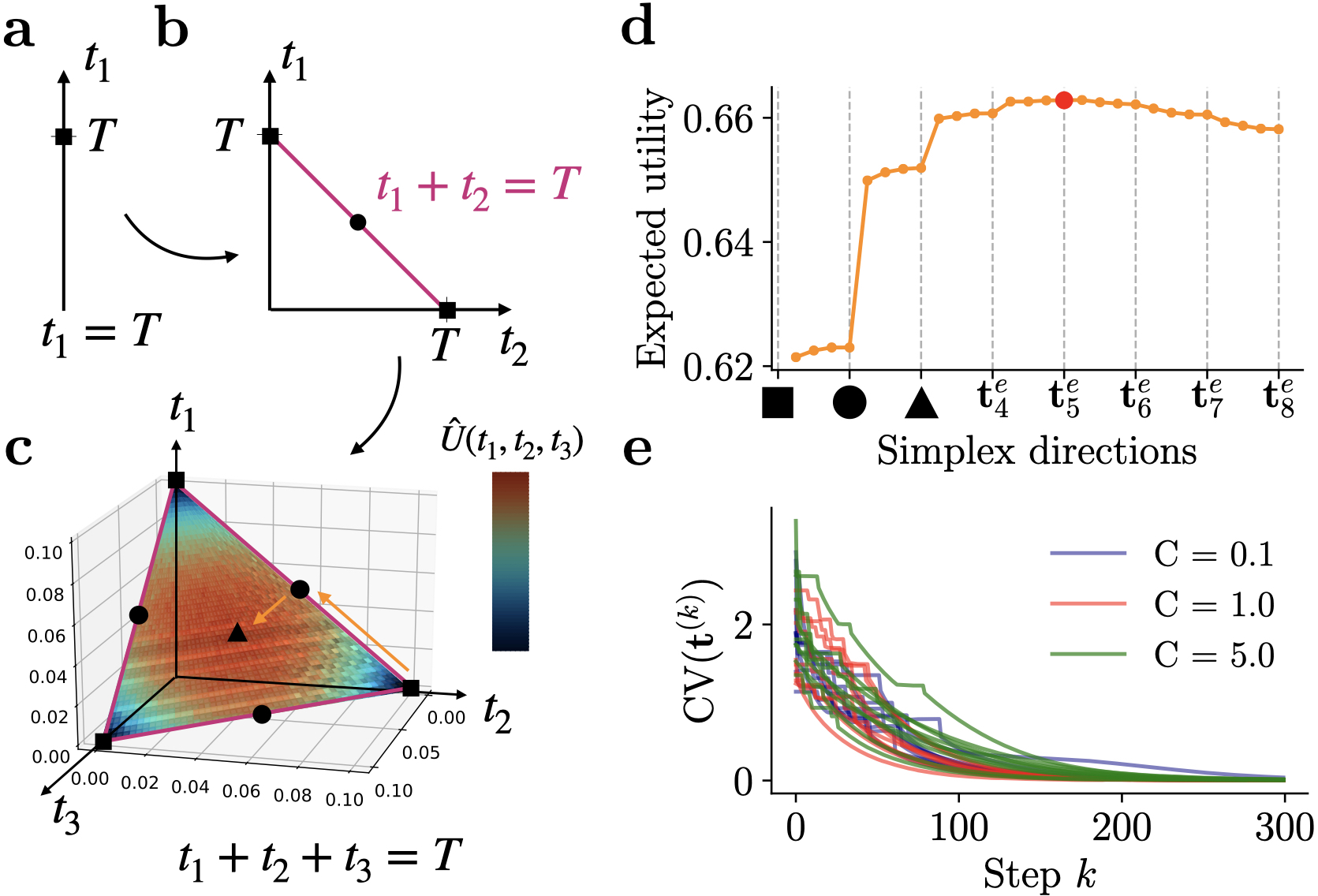} \newline
\\
 \caption{Even allocations correspond to critical points of utility lying at the center of $M$-simplices. 
 \textbf{(a)} In one dimension, there is only one point that complies with the constraint. 
 \textbf{(b)} For $M=2$ dimensions, constraints define a line segment or 1-simplex. The circle depicts the symmetric critical point $\mathbf{t}_2^e$. 
 \textbf{(c)} For $M=3$, constraints form a triangle or 2-simplex. The black triangle is the symmetric critical point $\mathbf{t}_3^e$.
 The colors at the extremes reflect the minimum and maximum utility reached in this simplex, which was computed with Monte Carlo simulations of Eq. (\ref{eq:utility_a}) for the Gaussian prior with $T = 0.1$, $\sigma = 1$, $\sigma_0 = 1$, $\mu_0 = 0.5$.  \textbf{(d)} Expected utility computed along directions that go orthogonally from $\mathbf{t}_{M}^e$ to $\mathbf{t}_{M+1}^e$ (as illustrated with orange arrows in panel \textbf{c}, same parameters). The red dot shows the maximum occurring at $\mathbf{t}_5^e$.
 \textbf{(e)}  Using the stochastic projected gradient ascent detailed in Sec. \ref{subsec:SGA}, we initialized the algorithm at random points (ten shown here) in a high-dimensional simplex and measured the coefficient of variation (CV) of the allocation vector at every step of the algorithm until convergence, for various values of capacity. Zero CV implies even allocation.}
 \label{fig:fig4}
 \end{figure}

In this more general setup, an even allocation corresponds to the symmetrical point in $M$ active dimensions given by $\mathbf{t}^e_M$, where $t^e_{M,i} =T/M$ for $i = 1,...,M$ (superscript reflects `even' allocation). 
As the expected utility in Eq. (\ref{eq:ut_arbitrary}) is symmetric under any permutation $t_j \leftrightarrow t_k$ for any $j$ and $k$, all its partial derivatives have to be equal at $\mathbf{t}_M^e$. Therefore, every even allocation for each $M$ corresponds to a critical point of the constrained optimization problem (see Sec. \ref{subsec:SGA}). 
 
We still need to characterize these critical points in order to show that the global maximum is indeed an even allocation. We first remember that the optimal number of active dimensions $M$ needs to be found, and thus it is useful to see how expected utility varies as a function of $M$. To do this, we note that any $M$-dimensional simplex is in fact the border of an ($M-1$)-dimensional simplex.
For example, for $M=2$, the constraints describe a line segment, or 1-simplex, where we have the symmetric critical point $\mathbf{t}^e_2=(T/2,T/2)$ (Fig. \ref{fig:fig4}\textbf{b}, black circle). However, the line $t_1 + t_2 = T$ is one of the 3 edges of the triangle, or 2-simplex (Fig. \ref{fig:fig4}\textbf{c}: pink lines are the edges of triangle), where in fact we have another symmetric critical point in its interior (black triangle).
With this, we can `visualize' the infinite-dimensional nature of this problem, since all critical points of the utility lie at the edges of a higher dimensional simplex.

To asses the landscape of expected utility in high-dimensional simplices, we can evaluate it at all symmetric critical points $\mathbf{t}^e_{M}$ and along directions that go orthogonally between them (Fig. \ref{fig:fig4}\textbf{c}, orange arrows). Thus, we devised a one-dimensional path that allows to continuously connect all symmetrical critical points, and applied it to the small capacity limit $C \ll 1$. As we move from the 1-simplex to higher dimensional simplices (as in Fig. \ref{fig:fig4}\textbf{c}), we find that first utility increases, reaching a maximum at the even allocation in $M=5$ dimensions, and then decreases (Fig. \ref{fig:fig4}\textbf{d}). Therefore, critical points $\mathbf{t}^e_2$, $\mathbf{t}^e_3$ and $\mathbf{t}^e_4$ are `saddle'-like points, as they are maxima in the interior of their corresponding simplex, and minima as one moves to the interior of the higher dimensional simplex. 

Although the above analysis suggests that the optimum lies at an even allocation point, it is still unclear whether there are other critical points that are asymmetrical and have a larger utility. To argue that the presence of non-symmetrical local optima is unlikely, we used a stochastic gradient projection method \cite{fletcher2013practical} that maximizes expected utility subject to the constraints, and applied it to the Gaussian prior case (see Sec. \ref{subsec:SGA} in Methods for details). Indeed, we find for various capacities that, regardless of the initial condition, i.e. random initial allocations, a maximum utility is attained when time is evenly divided (Fig. \ref{fig:fig4}\textbf{e}), and the global maxima coincide with the ones found in the previous sections.

\section{Discussion}

We have studied a model of multi-alternative decision making where an agent can allocate finite sampling resources to options and choose the best one amongst them. We found that the capacity of the agent depends on both the amount of sampling resources, i.e. time or precision, as well as on the discriminability of the options in the environment. As a function of capacity, optimal policies undergo an abrupt transition: at small capacity, allocating time to a handful of options is optimal; at large capacity, the number of options grows sub-linearly, well below the actual sampling capacity of the agent. Our results show that decision bottlenecks, such as option-narrowing, can arise from optimal policies in the face of uncertainty, and provide so far untested predictions on choice behaviors in multi-alternative decision making as a function of capacity.

Seemingly strict limits pervade cognition, from the so-called attentional bottleneck \cite{deutsch1963attention,treisman1969strategies,yantis1990locus}, over working memory \cite{miller1956magical,cowan2005capacity,luck2013visual,ma2014changing,brady2011review}, to executive control \cite{shenhav2017toward,norman1986attention,sleezer2016rule}. These limits might result from using scarce neuronal resources or from using them inefficiently. However, a likely alternative is that bottlenecks reflect strategies that make optimal use of limited but large resources. 
Indeed, past work has recognized that some apparent limits, most notably dual tasking bottlenecks \cite{fischer2015efficient,meyer1997computational}, could be the result of optimal allocation of finite resources to avoid overlap and interference between the different representations needed to solve the two tasks \cite{meyer1997computational,feng2014multitasking,zylberberg2011human}.  
Further, it has been recognized that the narrow focus of attention could be at the heart of solution to the the binding problem by integrating separate features into a coherent object \cite{treisman1998feature}, and thus its narrowness might reflect a function more than a limitation. Our work follows this line of argument and provides for the first time a quantitative account for why it is optimal for an agent to consider a handful of options in the face of uncertainty, well above two but well below 10. In addition, our results shed light on why people might ignore hundreds of accessible options and focus resources to a very small number of options \cite{hauser1990considerationset,iyengar2000choice,scheibehenne2010can}. Thus, some of the seemingly strict limits in decision making can be the result of optimal policies that favour depth versus breadth processing of the options. 

It has been long recognized that people often consider a small set of options while ignoring many others \cite{stigler1961economics,hauser1990considerationset,mehta2003price,payne1976task}. In the `consumer' literature this is explained by arguing
that small consideration sets are favored because they optimally balance the probability of finding a good option in the set with the search and mental costs incurred in adding new options to that set. These models thus assume that resources are not limited, but are costly. In contrast, the assumptions in our work do not explicitly tune the cost of sampling, but rather an implicit cost arises naturally from the strict capacity constraint, which depends intrinsically on the agent as well as extrinsically on the environment. A more fundamental distinction is that previous work did not focus on allocating resources intensively into the options, such that the only decision was whether to include an option into the set or not, without considering the amount of resources allocated to it. This distinction makes that problem drastically different than the tradeoffs of the BD dilemma considered here. This can explain why transitions of optimal policies as a function of agent's parameters have not been reported before. 

Previous work has characterized optimal BD tradeoffs in multi-alternative choices like the ones studied here, but by assuming that agents have a finite `discrete' capacity \cite{BDpnas}. Our assumption of a continuous resource (e.g. time) that can be infinitely divided has allowed us to uncover qualitatively novel optimal policies at small capacity. This is because a discrete small capacity can never produce numbers of sampled options above that capacity. Our modeling assumptions are also different and more in line with current theories of decision making based on accumulators of evidence \cite{gold2007neural,drugowitsch2012cost,moreno2010decision,ratcliff2004comparison} that trade accuracy over time.
Here we have not considered, however, a sequential process where accumulation of evidence can be stopped at any time, a topic that should be addressed in the future. In any event, any agent with finite capacity cannot avoid the problem of first deciding how many options to allocate capacity to, as dividing resources up to too small portions is clearly suboptimal, and thus BD tradeoffs as described above will be generally at play. Secondly, a sequential decision process can ensue after the first decision of how many options to sample, while we have assumed that allocated times cannot be reallocated on the fly. Although this will be as well a very relevant extension of our work, it is important to note that in many decisions it is actually hard, if not impossible, to reallocate already assigned resources, being neuronal, temporal or economical, and thus our framework more closely applies to those circumstances. 

Bounded rationality accounts \cite{simon1972theories,russell1991principles,gershman2015computational,griffiths2015rational} propose that cognition results from the finite limits of the nervous system from where it emerges. Our work follows this line of research in two ways. First, we propose that agents indeed have a finite sampling capacity that can be arbitrarily allocated to the available options. However, an important assumption in our work is that while the intrinsic resources of an agent might seem large, the interaction of the agent with the environment might render their effective decision-making capacity small. 
Therefore, capacity is not an absolute quantity that describes an agent, but a relative quantity that contextualizes the agents and characterizes how well they are suited to solve a given task in the world. 
An important contribution of our work is to show that optimal policies depend on effective capacity in a highly non-linear way, such that small-capacity agents would behave qualitatively different than large-capacity agents (or even the behavior of the same agent operating in different environments could be qualitatively different). This is clearly a prediction that can be tested with humans where time or other resources are constrained and varied on a trial by trial basis. 
Secondly, agents perform the allocation before feedback is received, which relates to a bounded-optimal agent that is optimized at `design'-time, which eliminates the paradox of perfect rationality by not letting the agent optimize their decisions at run-time \cite{russell1994boundedop}, an argument that further supports the validity and relevance of two-stage decisions. 

Another important result of our work is that evenly dividing time to a small set of options is optimal when they are initially indistinguishable. This optimal division of resources coincides with the $1/N$ heuristic rule \cite{heuristicgigerenzer2011} or equality heuristic \cite{messick1993equality}, which has proven to be implemented in human decision making and highly efficient as a portfolio strategy \cite{1/Ndemiguel2009}. In our case, the fact that options are drawn from the same prior (known to the agent) contributes to the optimality of the even allocation. Although the optimal allocation of non-identically distributed options is not addressed here, this heuristic can be efficient in such situations \cite{thorngate1980efficient}.
It is important to realize that the optimal low numbers of considered options have been found in the case where their values are not known in advance and come from the same distribution. If agents have strong preferences or have additional information about the expected values of the options, then the number of considered alternatives will be further reduced. This shows once again that a low number of considered options can hardly be taken as evidence of a decisional bottleneck and is more in line with an optimal tradeoff between breadth and depth. 

Finally, our results can have important implications for the optimal wiring of neural networks in the brain \cite{rushworth2011frontal,siegel2015cortical,vickery2011ubiquity,yoo2018economic}. First, as just few options should be considered at the same time, it is expected that only those would be encoded in different, albeit possibly overlapping, pools of neurons.
Thus, although models consisting of two or three pools that compete for dominance through mutual inhibition can be a sensible idea for binary and ternary decision making \cite{gold2007neural,cisek2010neural,roe2001multialternative,usher2001time,moreno2007noise,churchland2008decision,wang2008decision}, extrapolating this to many more options (e.g., larger than 10) by splitting neurons into corresponding pools of neurons would be hardly optimal. 
Our results are, in contrast, consistent with the opposite view that posits that a single pool of neurons is sufficient for decision making \cite{hayden2018neuronal}. In this framework, a single pool encodes just one of the available options, the one that is under the focus of attention. Previously attended options produce a background activity against which the current option is compared to, and other options fall outside the representation of the neural network \cite{hayden2018neuronal,krajbich2010visual,lim2011decision,redish2016vicarious,rich2016decoding}.
Thus, comparison and selection between options occurs through a temporal contrast, rather than through mutual inhibition between simultaneously encoded options. 
This model can be readily extrapolated to multiple many options, with the only dilemma of dividing time or precision into few or many options (like in Fig. 1), thus addressing the associated BD tradeoffs.
The debate of the one-pool versus several-pools models remains open \cite{hayden2018neuronal,ballesta2019economic}, but
electrophysiology experiments with many options should be able to arbitrate between the two hypotheses under the new computational constraints that we have identified here.

\section*{Acknowledgments}
This work is supported by the Howard Hughes Medical Institute (HHMI, ref 55008742), MINECO (Spain; BFU2017-85936-P) and ICREA Academia (2016) to R.M.-B, and MINECO/ESF (Spain; PRE2018-084757) to J.R.-R. J.R.-R. would like to thank Fred Callaway for helpful suggestions at early stages of this work.

\section*{Code availability}

All the numerical work performed to generate the various figures is available as documented Julia code along with a guided notebook at this public  \href{https://github.com/jorgeerrz/finite_time_allocation_paper}{GitHub repository}.

\bibliographystyle{unsrt}
\bibliography{references}

\section{Methods}\label{section:appendix}
Comments and mathematical proofs supporting claims in the manuscript.

\subsection{Posterior mean of drift is a monotonously increasing function of accumulated evidence}\label{subsec:expectedmu}
Here we prove that the posterior mean of the drift, which is a random variable with probability distribution defined by Bayes' rule, Eq. (\ref{eq:bayes}), is a monotonously increasing function of evidence $x$. We have seen that the expected value of a drift $\mu$ given the accumulated evidence $x$, for any option (and thus here dropping indices) is given by
\begin{equation}
    \hat{\mu}(x,t,\sigma,\theta) = \frac{\int \dd{\mu} \mu \mathcal{N}(x | \mu t, \sigma^2 t)p_\theta(\mu)}{p(x | t,\sigma,\theta)},
\end{equation}
where $p_\theta(\mu)$ is the prior probability of the drifts, with hyperparameters $\theta$ and $p(x | t,\sigma,\theta)$ is the marginalized probability distribution of the evidence. To know if $\hat{\mu}(x,t,\sigma,\theta)$ is an increasing function of $x$, we simply derive, and we expect the derivative to be always positive,
\begin{align*}
    \dv{\hat{\mu}}{x} = &\, \frac{p(x|t,\sigma,\theta)\int \dd{\mu} \mu \left(-\frac{x-\mu t}{\sigma^2t} \right)\mathcal{N}(x | \mu t, \sigma^2 t)p_\theta(\mu)}{p(x|t,\sigma,\theta)^2} \\ &- \frac{\int \dd{\mu} \mu \mathcal{N}(x | \mu t, \sigma^2 t)p_\theta(\mu)\int \dd{\mu} \left(-\frac{x-\mu t}{\sigma^2t} \right)\mathcal{N}(x | \mu t, \sigma^2 t)p_\theta(\mu)}{p(x|t,\sigma,\theta)^2} >0 \\
     \iff& \frac{1}{\sigma^2} \left(\mathbb{E}\left[\mu^2 | x, t, \sigma,\theta\right] - \mathbb{E}\left[\mu | x, t, \sigma,\theta\right]^2\right) = \frac{{\rm Var}\left[p(\mu|x,t,\sigma,\theta)\right]}{\sigma^2} > 0 ,
\end{align*}
where $$\mathbb{E}\left[\mu^n | x, t, \sigma,\theta\right] = \frac{\int \dd{\mu} \mu^n \mathcal{N}(x | \mu t, \sigma^2 t)p_\theta(\mu)}{\int \dd{\mu} \mathcal{N}(x | \mu t, \sigma^2 t)p_\theta(\mu)}.$$ Since the variance is the expected value of a positive quantity, then we conclude that the expected value of the drift is a monotonously increasing function of the observed accumulated evidence $x$ for any prior.

\subsection{Expected value of the drift in the small capacity limit}\label{subsec:evalmu}
Here we show that in the small capacity limit, the utility in Eq. (\ref{eq:utility_b}) can be written as in Eq. (\ref{eq:utility_small_C}) for any regular prior distribution. Our strategy is to study the limiting behaviors of the cumulative density function (described below in Sec. 4.3) and the posterior mean of the drift (detailed in this section) that appear in Eq. (\ref{eq:utility_b}) as $C = \frac{\sigma_0^2}{\sigma^2}T$ goes to zero. 

From Bayes's rule, Eq. (\ref{eq:bayes}), the posterior mean of the drift is given by
\begin{equation}\label{eq:expmu4.2}
    \hat{\mu}(x , t,\sigma,\theta) = \frac{1}{\sqrt{2\pi\sigma^2 t}}\frac{\int \dd{\mu} \mu \,\exp(-\frac{1}{2\sigma^2t}(\mu t-x)^2) p_\theta(\mu | \theta)}{p(x | t,\sigma,\theta)}.
\end{equation}
Let us focus on the numerator, which we will interpret as the expectation value of $\mu \,\exp(-\frac{1}{2\sigma^2t}(\mu t-x)^2)$ with respect to the prior. We assume the prior to be such that this expectation is finite for all $x$  and that all its moments are finite (e.g, Gaussian and uniform distributions). We define $z \equiv z(x) \equiv \frac{1}{\sqrt{\sigma^2t}}(x-\mu_0t)$ and $\mu_s \equiv \frac{1}{\sqrt{\sigma^2t}}(\mu t- \mu_0t)$, and by adding and subtracting $\mu_0t$ at the exponent, we can write the numerator in the above equation as

\begin{align*}
    \mathbb{E}_\theta\left[\mu \,\exp(-\frac{1}{2\sigma^2t}(\mu t-x)^2)\right] &= \mathbb{E}_\theta\left[\mu \,\exp(-\frac{1}{2\sigma^2t}(\mu t-\mu_0t + \mu_0t-x)^2)\right] \\
    & = \exp(-\frac{z^2}{2})\mathbb{E}_\theta\left[\mu\,\exp\left(z\mu_s - \frac{1}{2}\mu_s^2\right)\right] \\
    & = \exp(-\frac{z^2}{2})\Big\{\mathbb{E}_\theta\left[\mu_0\,\exp\left(z\mu_s - \frac{1}{2}\mu_s^2\right)\right] \\
    & + \mathbb{E}_\theta\left[\sqrt{\frac{\sigma^2}{t}}\mu_s\,\exp\left(z\mu_s - \frac{1}{2}\mu_s^2\right)\right]\Big\}.
\end{align*}

Next, we note that the exponential in the expectations is the generating function of the Hermite polynomials, and thus
\[
\exp(z\mu_s - \frac{1}{2}\mu_s^2) = \sum_{n=0}^\infty {\rm He}_n(z) \frac{\mu_s^n}{n!}.
\]
By replacing the exponential with the infinite series in the above expectation, Eq. (\ref{eq:expmu4.2}), we obtain
\begin{align*}
   p(x | t,\sigma,\theta)\hat{\mu}(x,t,\sigma,\theta) =& \frac{\exp(-\frac{z^2}{2})}{\sqrt{2\pi\sigma^2 t}} \left\{\mathbb{E}_\theta\left[\mu_0\,\sum_{n=0}^\infty {\rm He}_n(z) \frac{\mu_s^n}{n!}\right] 
     + \mathbb{E}_\theta\left[\sqrt{\frac{\sigma^2}{t}}\mu_s\,\sum_{n=0}^\infty {\rm He}_n(z) \frac{\mu_s^n}{n!}\right]\right\} \\
     =& \frac{\mathcal{N}(z | 0,1)}{\sqrt{\sigma^2t}} \left\{\sum_{n=0}^\infty  \frac{1}{n!}\mathbb{E}_\theta\left[\mu_s^n\right]\mu_0{\rm He}_n(z) + \sum_{n=0}^\infty  \frac{1}{n!}\mathbb{E}_\theta\left[\mu_s^{n+1}\right]\sqrt{\frac{\sigma^2}{t}}{\rm He}_{n}(z) \right\} \\
    =& \frac{\mathcal{N}(z | 0,1)}{\sqrt{\sigma^2t}} \left\{\sum_{n=0}^\infty  \frac{1}{n!}\mathbb{E}_\theta\left[\frac{(\mu-\mu_0)^n}{\sqrt{\sigma^2/t}^n}\right]\left(\mu_0{\rm He}_n(z) + \sqrt{\frac{\sigma^2}{t}}n{\rm He}_{n-1}(z)\right) \right\} \\
    =& \frac{\mathcal{N}(z | 0,1)}{\sqrt{\sigma^2t}} \Bigg\{\sum_{n=0}^\infty  \frac{1}{n!}\sqrt{\frac{C}{M}}^{\,n-1}\mathbb{E}_\theta\left[\frac{(\mu-\mu_0)^n}{\sigma_0^n}\right]\\
    &\quad\quad\quad\quad\quad
    \times\left(\sqrt{\frac{C}{M}}\mu_0{\rm He}_n(z) + \sigma_0n{\rm He}_{n-1}(z)\right) \Bigg\},
\end{align*}
where we have used that all the moments of the prior are finite and the sum is well defined. Note that to obtain the third line we have shifted the second index $ n + 1 \rightarrow n$ and used that the term $n{\rm He}_{n-1}(z)$ is zero for $n=0$.

We now insert the above series into the expression of utility in Eq. (\ref{eq:utility_b}) to obtain
\begin{align*}
    \hat{U}(t,\sigma,\theta) =& \int\dd{x}\, \dv{}{x}\left\{\left[F_x(x|t, \sigma, \theta)\right]^{M}\right\} \hat{\mu}(x , t,\sigma,\theta)  
    \\
    =& \int\dd{x}\, M \left[F_x(x|t, \sigma, \theta)\right]^{M-1} 
    \frac{\mathcal{N}(z | 0,1)}{\sqrt{\sigma^2t}} \\
    &
    \times \left\{\sum_{n=0}^\infty  \frac{1}{n!}\sqrt{\frac{C}{M}}^{\,n-1}\mathbb{E}_\theta\left[\frac{(\mu-\mu_0)^n}{\sigma_0^n}\right]\left(\sqrt{\frac{C}{M}}\mu_0{\rm He}_n(z) + \sigma_0n{\rm He}_{n-1}(z)\right) \right\} \\
    =& \int\dd{z}\, M \left[F_z(z|t, \sigma, \theta)\right]^{M-1} 
    \mathcal{N}(z | 0,1) 
    \left[\mu_0 +\sqrt{\frac{C}{M}} \sigma_0z\right]
    + \mathcal{O}(C),
\end{align*}
where in the second line it is implicit that $z$ depends on $x$, and in the last line we have made a linear transformation of variables from $x$ to $z=z(x)$. We also note that as the integral in the last line only involves polynomials in $z$ that are weighted by the standard normal (and by a cumulative, which is bounded to be in the range $[0,1]$), their integrals are finite, and thus we can truncate the series at the first leading order, which is order $\sqrt{C}$. It remains to see whether the cumulative density function $F_z(z|t, \sigma, \theta)$ contributes order $\sqrt{C}$ or larger, and we show below in Sec. (\ref{subsec:expexp})
that the former is actually true, such that $F_z(z|t, \sigma, \theta)=\frac{1}{2}\left[1 + \erf\left(\frac{z}{\sqrt{2}}\right)\right] + \mathcal{O}(C)$. With all this, we can approximate the utility up to order $\sqrt{C}$ as
\begin{equation}
    \hat{U}(C,M,\mu_0) = \mu_0 + \sigma_0\sqrt{C}\sqrt{M}\int_{-\infty}^\infty \dd{z} \left[\frac{1}{2}\left(1 + \erf\left(\frac{z}{\sqrt{2}}\right)\right)\right]^{M-1} \mathcal{N}(z | 0,1)\,z 
    +\mathcal{O}(C),
\end{equation}
which is identical to Eq. (\ref{eq:utility_small_C}).

\subsection{Distribution of evidence at small capacity limit}\label{subsec:expexp}

Here, we find an approximation to the marginalized probability distribution of the evidence at small capacity. From Bayes' rule and the law of the unconscious statistician,

\[
p(x | t, \sigma, \theta) = \int \dd{\mu} \mathcal{N}(x | \mu t,\sigma^2t)p_\theta(\mu | \theta) = \mathbb{E}_\theta\left[\mathcal{N}(x | \mu t,\sigma^2t)\right].
\]
 To compute this expectation, we follow the same procedure as in Sec. \ref{subsec:evalmu}. We define $z \equiv \frac{1}{\sqrt{\sigma^2t}}(x-\mu_0t)$ and $\mu_s \equiv \frac{1}{\sqrt{\sigma^2t}}(\mu t-\mu_0t)$ and add and subtract $\mu_0t$ at the exponent, to obtain
\begin{align*}
p(x | t, \sigma, \theta) &=
    \mathbb{E}_\theta\left[\exp(-\frac{1}{2\sigma^2t}(x-\mu t)^2)\right] \\
    &= \mathbb{E}_\theta\left[\exp(-\frac{1}{2\sigma^2t}(x - \mu_0t + \mu_0t - \mu t)^2)\right] \\
    & = \exp(-\frac{z^2}{2})\mathbb{E}_\theta\left[\exp\left(z\mu_s - \frac{1}{2}\mu_s^2\right)\right].
\end{align*}
Next, we again identify the exponential generating function of the Hermite polynomials,
\[
\exp(z\mu_s - \frac{1}{2}\mu_s^2) = \sum_{n=0}^\infty {\rm He}_n(z) \frac{\mu_s^n}{n!},
\]
and thus we obtain a series for the probability distribution of the evidence,
\begin{align*}
   p(x | t,\sigma,\theta)=& \frac{1}{\sqrt{2\pi\sigma^2 t}} \exp(-\frac{z^2}{2})\mathbb{E}_\theta\left[\sum_{n=0}^\infty {\rm He}_n(z) \frac{\mu_s^n}{n!}\right] \\
     =& \frac{\exp(-\frac{z^2}{2})}{\sqrt{2\pi\sigma^2t}} \sum_{n=0}^\infty  \frac{1}{n!}\mathbb{E}_\theta\left[\mu_s^n\right]{\rm He}_n(z)\\
    =& \frac{\exp(-\frac{z^2}{2})}{\sqrt{2\pi\sigma^2t}} \sum_{n=0}^\infty  \frac{1}{n!}\mathbb{E}_\theta\left[\frac{(\mu-\mu_0)^n}{\sqrt{\sigma^2/t}^n}\right]{\rm He}_n(z) \\
    =& \frac{\exp(-\frac{z^2}{2})}{\sqrt{2\pi\sigma^2t}} \sum_{n=0}^\infty  \frac{1}{n!}\sqrt{\frac{C}{M}}^{\,n}\mathbb{E}_\theta\left[\frac{(\mu-\mu_0)^n}{\sigma_0^n}\right]{\rm He}_n(z).
\end{align*}
We see that the leading order the distribution of the evidence is a normal distribution, while the order $\sqrt{C}$ is zero. Therefore, its cumulative in the variable $z=z(x)$ is, exactly, up to order $\sqrt{C}$, $F_z(z|t, \sigma, \theta)=\frac{1}{2}\left[1 + \erf\left(\frac{z}{\sqrt{2}}\right)\right] + \mathcal{O}(C)$. This expression has been used in Sec. (\ref{subsec:evalmu}).

\subsection{Asymptotic limit of relevant integral }\label{subsec:asymptotic}
In this subsection we want to obtain the asymptotic limit, $M \rightarrow \infty$, of the integral
\[
I(M) = \int_{-\infty}^\infty \dd{y} y\, \dv{}{y} \Phi^M(y),
\]
appearing in Eqs. (\ref{eq:utility_small_C}) and (\ref{eq:asymptoticGaussUtility}), where $\Phi(y)$ is the normal cumulative distribution function,
\[
\Phi(y) = \left(\frac{1}{2} + \frac{1}{2}\erf\left(\frac{y}{\sqrt{2}}\right)\right).
\]
Using Extreme Value Theory \cite{dehaanferreira2007extreme}, it can be shown that this cumulative distribution function $\Phi(y)$ belongs to the Gumbel class of the generalized extreme value distributions,
\[
\lim_{M \rightarrow \infty} \Phi^M(a_M y + b_M) = G(y), \]
where $G(y) = \exp(-\exp(-y))$ and 
\[
b_M = \left(2 \log(M) - \log(\log(M)) - \log(4\pi)\right)^{1/2} \quad \text{and} \quad a_M = 1/b_M .
\]
Using this result, then our integral develops quite easily,
\begin{align*}\label{eq:asymptotic_integral}
    I(M) &\rightarrow \int_{-\infty}^\infty \dd{y} y\, \dv{}{y} G\left(\frac{y - b_M}{a_M}\right) \nonumber \\
    & = \int_{-\infty}^\infty \dd{y} \left(\frac{y}{b_M} + b_M \right)\dv{}{y}G(y)\nonumber  \\
    & = \frac{1}{b_M}\int_{-\infty}^\infty \dd{y} y \exp(-y)\exp(-\exp(-y)) + b_M\int_{-\infty}^\infty \dd{y}\dv{}{y}G(y)\nonumber \\
    I(M \rightarrow \infty)& = \frac{\gamma}{b_M} + b_M ,
\end{align*}
where $\gamma \approx 0.577$ is Euler's constant.

\subsection{Expected utility for uniform prior}\label{subsec:uniform_prior}
For this choice of prior, drifts are all drawn independently and identically from a uniform probability distribution between zero and one. That is, 
\[
p(\mu_i) = \Theta(\mu_i)\Theta(1-\mu_i),
\] 
where $\Theta(x)$ is the Heaviside step function. We can substitute this prior into eq. (\ref{eq:bayes}) to obtain the posterior probability distribution for the drifts, 
\begin{equation*}
p(\mu_i | x_i, \sigma, t_i, \theta) =
 \begin{cases} 
      \frac{\mathcal{N}\left(\mu_i | \frac{x_i}{t_i},\frac{\sigma^2}{t_i}\right)}{\int_0^1 \mathcal{N}\left(\mu_i | \frac{x_i}{t_i},\frac{\sigma^2}{t_i}\right) \dd\mu_i}& \mu_i \in [0,1] \\
      0 & \text{otherwise} .
 \end{cases}
\end{equation*}
This will produce an expectation value for each drift,
\begin{equation}\label{eq:expectedmu_uniform}
   \hat{\mu}_i(x_i,t_i,\sigma) \equiv \mathbb{E}\left[\mu_i | x_i, t_i, \sigma \right] = \frac{x_i}{t_i} + \frac{\sigma}{\sqrt{2\pi t_i}}\frac{\exp\left(-\frac{x_i^2}{2\sigma^2t_i}\right)-\exp\left(-\frac{(x_i-t_i)^2}{2\sigma^2 t_i}\right)}{\frac{1}{2}\left[ \erf\left(\frac{x_i}{\sqrt{2\sigma^2t_i}}\right) -\erf\left(\frac{x_i-t_i}{\sqrt{2\sigma^2t_i}}\right)\right]},
\end{equation}
where the denominator is related to the probability distribution of the evidence $x_i$, which we can find by marginalizing over drifts,
\begin{align}\label{eq:probflat}
    p(x_i | t_i,\sigma) &= \int_0^1 \dd \mu \frac{1}{\sqrt{2\pi \sigma^2 t_i}}\exp\left(-\frac{1}{2\sigma^2t_i}(x_i-\mu_it_i)^2\right) \nonumber \\
    & = \frac{1}{2t_i}\left[ \erf\left(\frac{x_i}{\sqrt{2\sigma^2t_i}}\right) -\erf\left(\frac{x_i-t_i}{\sqrt{2\sigma^2t_i}}\right)\right].
\end{align}

We will use from now on the assumption of even time allocation, $t_i = t = \frac{T}{M}$ for all $i$. The cumulative  probability distribution for the evidence in eq. (\ref{eq:probflat})
is, integrating by parts, 
\begin{align*}
    F(x | t,\sigma) =& \int_{-\infty}^x p(x' |t,\sigma) \dd{x'} \nonumber \\
      = & \,\frac{1}{2}\left( 1 + \erf\left(\frac{x-t}{\sqrt{2\sigma^2t}}\right)\right) + \frac{x}{2t}\left(\erf\left(\frac{x}{\sqrt{2\sigma^2t}}\right)-\erf\left(\frac{x-t}{\sqrt{2\sigma^2t}}\right)\right) \nonumber \\ &+  \sqrt{\frac{\sigma^2}{2\pi t}}\left[\exp(-\frac{x^2}{2\sigma^2 t})- \exp(-\frac{(x-t)^2}{2\sigma^2 t})\right] \\
      & = \frac{1}{2}\left[ 1 + \erf\left(\frac{x-t}{\sqrt{2\sigma^2t}}\right)\right] + tp(x|t,\sigma)\hat{\mu}(x,t,\sigma),
\end{align*}
where in the last equality we have rewritten the solution in a convenient form.
Hence, the product of the expected value with the probability density can be rewritten in terms of the cumulative function, from the previous equation,
\[
\hat{\mu}(x,t,\sigma)p(x|t,\sigma) = \frac{1}{t}F(x|t,\sigma) -\frac{1}{2t}\left[ 1 + \erf\left(\frac{x-t}{\sqrt{2\sigma^2t}}\right)\right],
\]
and using eq. (\ref{eq:utility_b}) we get the expression for the utility, 
\begin{align}\label{eq:utility_flat}
    \hat{U}(M,t,\sigma) =& \frac{M}{t}\int_{-\infty}^{\infty} \dd x  \left[F(x| t,\sigma)\right]^{M-1}\left\{F(x| t,\sigma)- \frac{1}{2}\left[1 + \erf\left(\frac{x -t}{\sqrt{2\sigma^2t}}\right)\right] \right\}.
\end{align}

\subsection{Expected utility for Gaussian bimodal prior}\label{subsec:bimodal}

The expected utility for the bimodal Gaussian prior with modes $\mu_1$ and $\mu_2$, each with a variance $\sigma_0^2$, is quite similar to the unimodal, Eq. (\ref{eq:utilityG}), and follows the straightforward application of Eq. (\ref{eq:utility_b}). The probability distribution of the evidence marginalized over drifts is $p(x|t,\sigma,\theta) = \frac{1}{2}\mathcal{N}(x|\mu_1t,\sigma^2t + \sigma_0^2t^2) + \frac{1}{2}\mathcal{N}(x |\mu_2t,\sigma^2t + \sigma_0^2t^2)$. Therefore the cumulative is
\[
F(x|t,\sigma,\theta) = \frac{1}{2}\Phi(x|\mu_1t,\sigma^2t + \sigma_0^2t^2) + \frac{1}{2}\Phi(x|\mu_2t,\sigma^2t + \sigma_0^2t^2),
\]
where $\Phi(x|\mu_m,\sigma_m^2)$ is the normal cumulative distribution for one mode.
However, the expected value of the drift is a bit more involved, since the posterior distribution over drifts takes a different form,
\[
p(\mu|x,t,\sigma,\theta) = \frac{\sigma_t}{\sqrt{2\pi\sigma^2t\sigma_0^2} p(x|t,\sigma,\theta)} \sum_i \frac{1}{2} \mathcal{N}(\mu | \hat{\mu}_i,\sigma_t^2)\exp\left(-\frac{1}{2(\sigma_0^2t^2+\sigma^2 t)}\left(\mu_it - x\right)^2\right),
\]
where $1/\sigma_t^2 = t/\sigma^2 + 1/\sigma_0^2$ and 
\[
\hat{\mu}_i= \frac{\sigma^2_t}{\sigma_0^2}\mu_i + \frac{\sigma^2_t}{\sigma^2}x.
\]
Consequently, the expected value will be
\[
\hat{\mu}(x,t,\sigma,\theta) = \frac{1}{\sqrt{2\pi(\sigma_0^2t^2 + \sigma^2t)}}\frac{1}{ p(x|t,\sigma,\theta)} \sum_i \frac{\hat{\mu}_i}{2}\exp\left(-\frac{1}{2(\sigma_0^2t^2+\sigma^2 t)}\left(\mu_i t - x\right)^2\right).
\]
Then, expected utility is
\begin{equation}\label{eq:utility_bimodal}
    \hat{U}(M,t,\sigma,\theta) =  M\int_{-\infty}^{\infty}\dd{x}F(x|t,\sigma,\theta)^{M-1}\sum_i \frac{\hat{\mu}_i}{2}\mathcal{N}\left(x |\mu_i t,\sigma_0^2t^2 + \sigma^2t\right).
\end{equation}
This expression is numerically integrated and used in Fig. \ref{fig:fig3}.

\subsection{Stochastic gradient ascent method for Gaussian prior }\label{subsec:SGA}

To maximize utility, Eq. (\ref{eq:ut_arbitrary}), under the time constraint, we can make use of unconstrained optimization through Lagrangian multipliers. We construct the Lagrangian given by
\begin{equation}\label{eq:lagrangian}
L(\mathbf{t},\theta, \lambda)  = \hat{U}(\mathbf{t},\sigma,\theta) + \lambda h(\mathbf{t}) + \boldsymbol{\mu}\cdot \mathbf{g}(\mathbf{t}),
\end{equation}
where $h(\mathbf{t}) = \sum_{i=1}^M t_i - T$ is the equality constraint that defines the hyperplane and $0 \le g_i(\mathbf{t}) = t_i$ is the inequality constraint forcing all times $i$ to be non-negative and thus defining the simplex. The quantities $\lambda$ and $\boldsymbol{\mu}$ are the Lagrangian multipliers. In other words, maximizing utility, Eq. (\ref{eq:ut_arbitrary}), subject to the initial constraints can be done by optimizing the Lagrangian, Eq. (\ref{eq:lagrangian}), with respect to $\mathbf{t}$, $\lambda$ and $\boldsymbol{\mu}$ subject to Karush-Kuhn-Tucker conditions \cite{bishop2006pattern}
\begin{subequations}\label{eq:KKTconditions}
\begin{align}
    g_i(\mathbf{t}) &\ge 0, \quad \text{for all }  i \\
    \mu_j &\ge 0, \quad \text{for all }  j\\
    \boldsymbol{\mu} \cdot \mathbf{g}(\mathbf{t}) &= 0
\end{align}
\end{subequations}
We notice that the first two conditions imply that the third can be rewritten as $\mu_it_i =0$ for all $i$.
By optimizing the Lagrangian, Eq. (\ref{eq:lagrangian}), we obtain the following system of equations 
\begin{equation}\label{eq:optLagrangian}
    \nabla_{\mathbf{t}}\, \hat{U}(\mathbf{t}^*,\sigma,\theta) + \lambda^*\mathbf{1} + \boldsymbol{\mu}^* = \mathbf{0},
\end{equation}
where $\mathbf{t}^*,\lambda^*,\boldsymbol{\mu}^*$ denote the critical points of the Lagrangian.
We note that the symmetrical point in $M$ active dimensions, denoted by $\mathbf{t}^e_M$, where $t^e_{M,i} =T/M$ for $i = 1,...,M$, is a critical point of the Lagrangian. This is because the partial derivatives with respect to the utility have to be equal at $\mathbf{t}_M^e$, and since this point lies in the interior of the $(M-1)$-simplex, the $\boldsymbol{\mu}^e_i = 0$ for $i=1,\ldots,M$. Therefore $\mathbf{t}_M^e$ complies with Eq. (\ref{eq:optLagrangian}) and is indeed a critical point.

Next, we detail the gradient ascent method used to obtain Fig. \ref{fig:fig4}\textbf{e}. As explained above, we want to optimize utility, Eq. (\ref{eq:ut_arbitrary}), subject to a set of equality, Eq. (\ref{eq:T-constraint}), and inequality constraints, $t_i \ge 0$, as described in section ``Even allocation is optimal" of the Results. As all our constraints are linear, we can make use of the gradient projection method \cite{fletcher2013practical}. In this case, we want to obtain the gradient of utility in Eq. (\ref{eq:utilityG}) and project it in the $(M-1)$-simplex such that the capacity constraint in Eq. (\ref{eq:T-constraint}) is satisfied.
Due to the linear capacity equality constraint, this projection is simply given by the linear operator
\[
\Pi = \mathbf{Id}_{M\times M} - \frac{1}{M}\mathbf{1}_{M\times M}
\]
where $\mathbf{Id}_{M\times M}$ is the $M\times M$ identity matrix and $\mathbf{1}_{M\times M}$ is an $M\times M$ matrix full of ones. Therefore, we can maximize utility by updating $\mathbf{t}^{(k)}$ appropriately,
\begin{equation}\label{eq:update}
    \mathbf{t}^{(k+1)} = \mathbf{t}^{(k)} + \eta \Pi \left(\nabla_{\mathbf{t}}\,\hat{U}(\mathbf{t},\sigma,\theta)\right),
\end{equation}
where $\eta=10^{-1}\,T$ is the default step size, $k$ is the iteration number, and $\theta$ corresponds to the parameters of the Gaussian prior. The utility for an arbitrary time allocation $\mathbf{t}$ for the Gaussian prior case is, using Eq. (\ref{eq:ut_arbitrary}), 
\begin{align}\label{eq:utility_arbitrary_Gaussian}
    \mathbf{\hat{U}}(\mathbf{t},\sigma,\theta) 
    &= \mu_0 + \sigma_0\sum_{i=1}^N \int_{-\infty}^{\infty} \dd{y}\,y\,  \frac{\exp(-\frac{y^2}{2\sigma_i^2})}{\sqrt{2\pi\sigma_i^2}}\prod_{j\neq i}\left\{\frac{1}{2}\left[1 + \erf\left(\frac{y}{\sqrt{2}\sigma_j}\right)\right]\right\} ,
\end{align}
where $\sigma_i^2 = \frac{\sigma_0^2 t_i}{\sigma_0^2 t_i + \sigma^2}$. We can therefore compute the derivative of the previous equation with respect to all components $t_i$ and numerically integrate the expression that results.

In addition to the linear capacity constraint, we have to enforce the inequality constraints as well, i.e. $t_i \ge 0$, which we do by utilizing an active set of constraints. To implement it, we start in a relatively high-dimensional ($M-1$)-simplex, choosing $M$ to be $2M^*$, where $M^*$ is the optimal number of accumulators to sample in the even sampling case (which is estimated before through exploration, see main text). If and whenever any of the components of $\mathbf{t}^{k+1}$ derived from Eq. (\ref{eq:update}) is approaching a border ($t_i^{(k+1)}\approx \tau$ for some $i$ and small $\tau$), the step size decreases until the component effectively reaches zero. In such a case, this dimension is added to the active constraints set (we inactivate the dimension), thus downgrading the simplex to a lower dimension. In this way, our algorithm only reduces the initial dimension of the simplex and never extends it. To initially activate the $2M^*$ dimensions, for any random initial condition $\mathbf{t}_0$, we make sure that all the components are greater than our threshold $t_{0,i} > \tau$ for all $i = 1,...,2M^*$.

Finally, in order to avoid potentially getting trapped in local maxima, we add noise at every iteration as follows. At every step $k$ of Eq. (\ref{eq:update}), and with probability $\epsilon=0.1$, we push the $t^{(k)}_i$ of a randomly chosen dimension $i$ by a magnitude $\delta = 10^{-3}T$ and pull the $t^{(k)}_j$ of another random dimension $j$ in the opposite direction with the same amount in order to stay in the appropriate simplex.

\end{document}